\documentclass[twocolumn,aps,apl,showpacs,superscriptaddress,10pt]{revtex4}
\usepackage{amsmath}
\usepackage{amsfonts}
\usepackage{amssymb}
\usepackage{graphicx}
\usepackage[colorlinks,
            linkcolor=blue,
            anchorcolor=blue,
            citecolor=blue,
            urlcolor=blue
            ]{hyperref}

\begin{document}
\title{Exotic superfluid of trapped Fermi gases with spin-orbit coupling in
dimensional crossover}
\author{Jing Zhou}
\affiliation{Department of Science, Chongqing University of Posts and Telecommunications,Chongqing 400065, China }
\affiliation{Beijing National Laboratory for Condensed Matter Physics, Institute of
Physics, Chinese Academy of Sciences, Beijing 100190, China }
\author{Cheng Shi}
\affiliation{Department of Science, Chongqing University of Posts and Telecommunications,Chongqing 400065, China }
\author{Xiang-Fa Zhou}
\affiliation{Key Laboratory of Quantum Information, University of Science and Technology
of China, CAS, Hefei, Anhui, 230026, People's Republic of China}
\author{Lin Wen}
\affiliation{College of Physics and Electronic Engineering, Chongqing Normal University,
Chongqing, 401331, China}
\author{Peng Chen}
\affiliation{Department of Science, Chongqing University of Posts and Telecommunications,Chongqing 400065, China }
\author{Deng-Feng Li}
\affiliation{Department of Science, Chongqing University of Posts and Telecommunications,Chongqing 400065, China }

\begin{abstract}
We have studied the exotic superfluid phases of degenerated Fermi gases with
spin-orbit coupling in a mixed dimensional system, where the motion of atoms
are free in the $\hat{x}$-direction and the tunneling between nearest tubes in the $%
\hat{y}$-direction is permitted. Using the mean-field method, we obtain the
phase diagrams of the system during the dimensional crossover between quasi
one dimension to quasi two dimension. We find the existence of the
topological state and Majorana edge mode in the weak tunneling case, and a
rich phase diagram including two kinds of nodal superfluid phase and gapped
superfluid phase in the opposite case. Our results show that topological
pairing is favoured in quasi one dimension while nodal pairing state is
favoured in quasi two dimension.
\end{abstract}

\maketitle


\section{introduction}
Spin-orbit coupling (SOC) has received tremendous attention
due to its rich physics. For
instance, it gives rise to novel transport properties in semiconductor
materials \cite{niu,xie,Hsieh}, and plays a crucial role in the formation of
many important phenomena including topological insulator \cite{kane1},
topological conductor \cite{Qi}, and quantum spin Hall effect \cite{kane}.
In recent years, with the help of Raman laser coupling, SOC has been
implemented experimentally \cite{Lin,Wang,Zwierlein} in ultra-cold neutral
atomic system and has attracted considerable attentions both theoretically
and experimentally \cite%
{zhang1,zhang2,zhang3,zhang4,Yu,Gong,Zhou,Cheng,He1,He2,He3,Shenoy,Gong2,Hu0,Hu1,Hu2,Hu3,Hu4,Hu5,Hu6}%
. Due to its highly flexibility, cold atom system is considered as an ideal
platform for the quantum simulation of interesting physics models, ranging
from nuclear physics \cite{higgs1,higgs2} to condensed matter physics \cite%
{bloch2,duan}. This provides a great opportunity to investigate such novel
SOC physics in a highly controllable way.

Fulde-Ferrell-Larkin-Ovchinnikov (FFLO) state first proposed
in the 1960s \cite{FF1,FF2} is an exotic superfluid pairing state
with finite momentum compared with the conventional superconductor with zero
momentum pairing. In general, FFLO state emerges with mismatched fermi
surface of imbalanced Fermi gas. Therefore, it has been viewed as one of the
most interesting superconduction phases, and is studied in various field,
ranging from heavy fermions \cite{wu1,wu2,wu3}, organic superconductor \cite%
{org1} to ultra-cold atomic gases \cite{fflo1,fflo2,fflo3,fflo4,fflo5,fflo6}%
. However, due to the small pairing phase space in 2D and 3D, only phase
separation is observed in experiments rather than FFLO state. The difficulty
is avoided in 1D system, where the fermi surface is point, and the pairing
phase space isn't affected by the spin imbalanced. But the large quantum fluctuation
in 1D might suppress the long-range order. With spin-orbit coupling and
Zeeman field, the spatial inversion symmetry is broken and paring on a
single Fermi surface is passible. Theoretical investigation indicates that
FFLO state is favoured in this system with nontrivial topology \cite{Qu,Yi,Liu,Cao}.

In this paper, we consider an array of Fermi tubes with one dimension SOC.
The atoms can move freely along the $\hat{x}$-direction while nearest tube
tunneling is allowed in our model \cite{Sun}. The tunneling can be adjusted over a
wide range by tuning the optical lattice height in the $\hat{y}$-direction so
that we can realize the crossover from quasi one dimensional system to
quasi two dimensional ones. With different tunneling strength, we obtain
the mean-field ground state, and confirm that the FFLO state is hosted in such dimensional crossover system.
Furthermore, we show topological FFLO state is more stable with smaller tunneling, which can be
verified by the appearance of typical Majorana zero mode. To obtain the
phase transition point between topological FFLO state and normal FFLO state,
we use two coupled tubes model to analyze the phase transition and find the
critical tunneling. It indicates that topological FFLO state exists not only in 1D
\cite{Hu0,Hu4}, but also in quasi 1D. For quasi two dimensional system, we find the system
supports novel gapless superfluid phases which can be classified by the
number of gapless points in the excitation spectra. The influence of
transverse Zeeman field, interaction strength, and tunneling between
nearest tubes on these novel phases are also studied. Detailed calculation
shows that the presence of transverse Zeeman field eradicates the zero
center-of-mass momentum pairing, and stabilizes nodal FFLO state with finite
center-of-mass momentum pairing along anisotropic SOC direction. We also give the
phase diagrams with different interactions and tunneling and show the interaction
effect and how the
crossover physics can be explored by tuning the tunneling.

The rest of this paper is organized as follows: in the next section (Sec.$%
\text{II}$), we introduce our model and detailed calculation method based on
the standard Bogoliubov-de Gennes (BdG) theory. In Sec.$\text{III}$, we discuss
the phase diagram of quasi one dimensional system with small inter-tube
tunneling. Two coupled tubes model is used to analyze the critical
tunneling between topological FFLO state and normal FFLO state, where
topological properties can be verified by the presence of Majorana edge
modes. In Sec. $\text{IV}$, we investigate the phase diagram of
quasi two dimensional system. The effects of transverse Zeeman field,
interaction and tunneling are studied to capture the properties of the
exotic nodal superfluid states, where the crossover physics can also be
clearly manifested. Finally, we summarize the results and draw our
conclusion in Sec. $\text{V}$.

\section{MODEL HAMILTONIAN AND CALCULATION METHOD}

\begin{figure}[ptb]
\includegraphics[width=7.5cm]{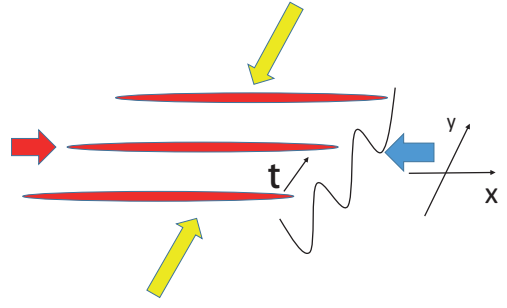}\newline
\caption{Illustration of the model in an array of coupled tubes in
two-dimensional plane, which can realize dimensional crossover.}
\label{f1}
\end{figure}

We start by considering two-component Fermi gas with SOC and Zeeman field in
an array of tubes. As shown in Fig.\ref{f1}, the red tubes are parallel to the
$\hat{x}$-direction. Two counter-propagating laser lights along the $\hat{y}$-direction construct
one dimension optical lattice in this direction. The tunneling between nearest
tubes is permitted in the model. The type of SOC is induced by another
two counter-propagating Raman laser lights along $\hat{x}$-direction.
The Hamiltonian can be written as%
\begin{equation}
H-\sum_{\sigma}\mu_{\sigma}N_{\sigma}=H_{0}+H_{SOC}+H_{Zee}+H_{int},
\end{equation}
with
\begin{align*}
H_{0} & =\sum_{\sigma,k_{x},j}\varepsilon_{k}C_{k_{x}j\sigma}^{\dagger
}C_{k_{x}j\sigma}-t\sum_{\sigma,k_{x},\langle i,j\rangle}C_{k_{x}i\sigma
}^{\dagger}C_{k_{x}j\sigma}+h.c., \\
H_{SOC} & =\sum_{k_{x},j}(\alpha
k_{x}C_{k_{x}j\uparrow}^{\dagger}C_{k_{x}j\uparrow}-\alpha
k_{x}C_{k_{x}j\downarrow}^{\dagger}C_{k_{x}j\downarrow}), \\
H_{Zee} &
=\sum_{k_{x},j}h_{x}(C_{k_{x}j\uparrow}^{\dagger}C_{k_{x}j%
\downarrow}+C_{k_{x}j\downarrow}^{\dagger}C_{k_{x}j\uparrow}), \\
& -h(C_{k_{x}j\uparrow}^{\dagger}C_{k_{x}j\uparrow}-C_{k_{x}j\downarrow
}^{\dagger}C_{k_{x}j\downarrow}), \\
H_{int} & =g\sum_{\mathbf{k,k}^{^{\prime}},\mathbf{Q}}C_{\mathbf{k+Q/2}%
\uparrow}^{\dagger}C_{\mathbf{-k+Q/2}\downarrow}^{\dagger}C_{\mathbf{-k}%
^{^{\prime}}\mathbf{+Q/2}\downarrow}C_{\mathbf{k}^{^{\prime}}\mathbf{+Q/2}%
\uparrow},
\end{align*}
where $\varepsilon_{k}=\frac{\hbar^{2}k_{x}^{2}}{2m}-\mu$, $C_{k_{x}\sigma j}$
($C_{k_{x}\sigma j}^{\dagger}$) is the annihilation (creation) operator for
the hyperfine spin state $\sigma$ in the $j$-th tube. $m$ is the atom mass, $%
\alpha$ denotes the strength of the SOC, $t$ is the near-rest tunneling
between different tubes, and $h.c.$ represents the Hermitian conjugate.
The Zeeman field contains two terms: in-plane ($h_{x}$) and out-of-plane ($h$%
) magnetic fields, which are included in the current experimental scheme. $g$
is the bare strength of s-wave attractive interaction.

Throughout the paper, we only focus on the zero temperature properties of
the system. The order parameter of FF pairing state is defined as $\Delta _{%
\mathbf{Q}}=g\sum_{\mathbf{k}}\langle C_{\mathbf{-k+Q/2}\downarrow }C_{%
\mathbf{k+Q/2}\uparrow}\rangle$. Within mean field approximation, we can rewrite the Hamiltonian
based on the Nanbu spinor base $\Psi_{\mathbf{k}}=[C_{\mathbf{k+Q/2}\uparrow
},C_{\mathbf{k+Q/2}\downarrow},C_{\mathbf{-k+Q/2}\uparrow}^{\dagger },C_{%
\mathbf{-k+Q/2}\downarrow}^{\dagger}]^{T}$ as follows
\begin{equation}
\hat{H}=\frac{1}{2}\sum_{\mathbf{k}}\hat{\Psi}_{\mathbf{k}}^{\dagger}H_{eff}\hat{\Psi}_{\mathbf{k}}+\sum_{\mathbf{k}}\zeta_{-\mathbf{k}+\mathbf{Q}/2}-\frac{\Delta_{\mathbf{Q}}^2}{g}
\end{equation}
with the BdG Hamiltonian%
\begin{align}
H_{eff} & =\left(
\begin{array}{cccc}
\lambda_{\mathbf{k+Q/2}}^{+} & h_{x} & 0 & \Delta_{\mathbf{Q}} \\
h_{x} & \lambda_{\mathbf{k+Q/2}}^{-} & -\Delta_{\mathbf{Q}} & 0 \\
0 & -\Delta_{\mathbf{Q}} & -\lambda_{-\mathbf{k+Q/2}}^{+} & -h_{x} \\
\Delta_{\mathbf{Q}} & 0 & -h_{x} & -\lambda_{-\mathbf{k+Q/2}}^{-}%
\end{array}
\right),
\end{align}
where $\lambda_{\mathbf{k}}^{\pm}=\hbar^{2}k_{x}^{2}/2m-2t(\cos(k_{y})-1)%
\pm(\alpha k_{x}-h)$, and $\zeta_{k}=\hbar^{2}k_{x}^{2}/2m-2t(\cos
(k_{y})-1)-\mu$. Since there is no constrain in the $\hat{x}$-direction, $%
k_{x}$ still remains to be a good quantum number, while $k_{y}$ is confined
in the first Brillouin $|k_{y}|\leq\pi$. The tunneling $t$ is highly
controllable by changing the lattice height in the $\hat{y}$-direction, which
allow us to investigate the crossover physics from 1D to 2D.

In order to get the zero temperature phase diagram, we straightforwardly
diagonalize the BdG Hamiltonian $H_{eff}$ and calculate the thermodynamic
potential $\Omega =-\frac{1}{\beta }\log \mathrm{Tr}e^{-\beta (H-\mu N)}$.
For zero temperature, the thermodynamic potential is:%
\begin{equation}
\Omega =\sum_{\mathbf{k}}\zeta _{-\mathbf{k+Q/2}}+\sum_{\mathbf{k,\nu ,\eta }%
}\Theta (-E_{\mathbf{k,\eta }}^{\nu })E_{\mathbf{k,\eta }}^{\nu }-\frac{{%
|\Delta _{\mathbf{Q}}|^{2}}}{{g}},
\end{equation}%
where the quasi-particle (hole) dispersion $E_{\mathbf{k,\eta }}^{\nu }$ $%
(\eta ,\nu =\pm )$ are the eigenvalues of the BdG Hamiltonian. And $%
\Theta $ is the Heaviside step function. For sake of simplicity, we can
assume the order parameter $\Delta _{\mathbf{Q}}$ to be real throughout the
whole work. In this model, due to the complex $4\times 4$ matrix $H_{eff}$, the analytical expression of quasi-particle dispersion $E^{\nu}_{k,\eta}$ are too complicated to be presented here,
so we will perform our calculation numerically. In general, the ground state can be obtained from the saddle point equations $\partial
\Omega /\partial \Delta =0$, $\partial \Omega/\partial \mathbf{Q}=0$ and $\partial \Omega/\partial \mu=-n$. However, Due to the competition between pairing and the Zeeman
field, the thermodynamic potential shows double well structure. We may get
metastable state from the gap equation. In order to find the true ground
state, we numerically search for the global minimum of the thermodynamic
potential to obtain the order parameter $\Delta _{\mathbf{Q}}$ as well as
the finite mass momentum $\mathbf{Q}$. We note that in our work, the phase
diagram is obtained by fixing the chemical potential throughout the whole
system. For a trapped gas system, the local chemical potential is defined as
$\mu (\mathbf{r})=\mu -V(\mathbf{r})$ , which is known as the local density
approximation (LDA). In principle, based on our calculation, we can sweep
chemical potential to obtain the phase structure in a trapped system. The
total particle number can be evaluated by integrating the local particle
density from the trap center to the edge.

\section{QUASI ONE DIMENSIONAL RESULT}

\subsection{PHASE DIAGRAM}

For the fixed chemical potential, the paring
order parameter is obtained by searching the global minimum of the
thermodynamic potential. In the calculation, we find with not large enough $h$,
FF state with finite paring momentum $\mathbf{Q}=Q\widehat{x}$ is stable
in a fairly large parameter region. We map out the phase diagram in $%
\alpha-\mu$ plane, where a topological FF phase (tFF) is always sandwiched
between a normal FF state (gFF) and normal state (N) without pairing.

The tFF state can be understood from the symmetry of the BdG Hamiltonian
\begin{equation}
H_{BdG}(\mathbf{k})=\left(
\begin{array}{cccc}
\lambda_{\mathbf{k+Q/2}}^{+} & h_{x} & 0 & \Delta_{\mathbf{Q}} \\
h_{x} & \lambda_{\mathbf{k+Q/2}}^{-} & -\Delta_{\mathbf{Q}} & 0 \\
0 & -\Delta_{\mathbf{Q}} & -\lambda_{-\mathbf{k+Q/2}}^{+} & -h_{x} \\
\Delta_{\mathbf{Q}} & 0 & -h_{x} & -\lambda_{-\mathbf{k+Q/2}}^{-}%
\end{array}
\right) .
\end{equation}
We note that this Hamiltonian processes a particle-hole symmetry as $\Xi
H_{BdG}\Xi^{-1}=\Lambda H_{BdG}^{\ast}(\mathbf{k})\Lambda=-H_{BdG}(-\mathbf{k%
})$. Here, $\Xi=\Lambda K$, $\Lambda=(\sigma_{y}\tau_{y})$ and $K$ is the
complex conjugation operator. $\sigma_{x,y,z}$ are the Pauli-spin matrices
and $\tau_{x,y,z}$ are the Nanbu particle-hole matrices. In this case, the
topological character is verified by the topological number - Pfaffian
invariant Q \cite{Pf}, which can in principle be computed in terms of the
eigenvalues of $H_{BdG}(\mathbf{k})$. The topological phase corresponds to $%
Q=-1$, while the topological trivial state corresponds to $Q=1$. In general,
the topological nature is protected by the gap in the bulk quasi-particle
excitation. The phase boundary between tFF state and gFF state can be
determined by the gap condition. According to the symmetry, the gap closes
at a single point $\mathbf{k}=(0,0)$. With simple matrix determinate
calculation, we find the energy gap close and reopen condition is
\begin{equation}
h^{2}=\Delta_{\mathbf{Q}}^{2}-h_{x}^{2}+\alpha hQ_{x}-\frac{1}{4}\alpha
^{2}Q_{x}^{2}+(\mu-\frac{Q_{x}^{2}}{4})^{2}.
\end{equation}
In the special case that both the in-plane Zeeman field $h_{x}$ and the paring
momentum $Q_{x}$ equals to zero, the above formula reduces to the standard
condition for BCS topological superfluid $h^{2}=\Delta^{2}+\mu^{2}$ \cite%
{Gong}.

\begin{figure}[ptb]
\includegraphics[width=8.4cm]{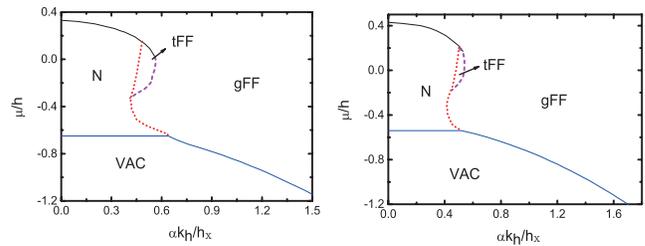}\newline
\caption{The phase diagram in the $\protect\alpha-\protect\mu$ plane with
different tunneling $t$ (Left: $t=0.02$; Right: $t=0.08$ ). The first order
phase transition is shown in black solid line while the topological phase
transition in dashed burgundy curve. The red dotted line marks the $%
\Delta/h_{x}=10^{-3}$ threshold. The in-plane Zeeman field is taken as the
energy unit, while the unit of momentum $k_{h}$ is defined through $%
\hbar^{2}k_{h}^{2}/2m=h_{x}$. The other parameters used in this plot are:$%
g/h_{x}=-0.5,h/h_{x}=0.2$. }
\label{f2}
\end{figure}

In Fig. \ref{f2}, we show the phase diagrams of quasi-1D system, where the
first order phase transition is shown in black solid line while the
topological phase transition is shown in dashed burgundy curve. The red dotted line
marks the $\Delta/h_{x}=10^{-3}$ threshold. The phase boundary between gFF
state and normal state is determined from the double well structure of the
thermodynamic potential. The tFF state in the phase diagram is verified by
the topological invariant Pfaffian and the topological condition above.
Comparing the two plots in Fig. \ref{f2}, it is obvious that tFF state
region is much larger with weaker tunneling $t$. This means that tFF state
of our system is much more stable in quasi one dimensional geometry than in
quasi two dimensional one.

\subsection{TWO COUPLED TUBES MODEL}

\begin{figure}[ptb]
\includegraphics[width=8cm]{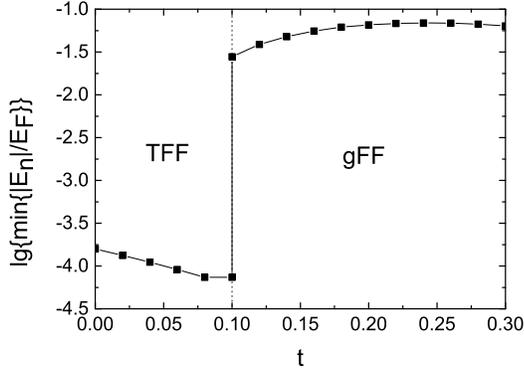}\newline
\caption{Phase diagram for a given spin-orbit coupling $\protect\alpha %
k_{F}/E_{F}=0.6$ of two coupled tubes model, is determined from the real
space BdG calculation. The Bogoliubov excitation spectra is calculated by
diagonalizing a large matrix in the real space. As tunneling increases, a phase
transition occurs, and the system evolves from a tFF state to a gFF state. The
critical tunneling $t=0.1$. The other parameter is set as: $%
g=-0.3,h=0.2,hx=0.2,\protect\alpha k_{F}/E_{F}=0.6$. The energy unit $E_{F}$
is determined by the total particle number. }
\label{f3}
\end{figure}

To investigate the effect of tunneling on phase transition, we
further consider a simplified toy model with only two coupling tubes. Within
mean-field approximation, we calculate the elementary excitations using the
standard Bogoliubov-de Gennes equation as
\begin{equation}
H_{BdG}(x)\Psi(x)=E_{n}\Psi(x),
\end{equation}
where $\Psi(x)=[u_{n,1}^{\uparrow},u_{n,1}^{\downarrow},u_{n,2}^{\uparrow
},u_{n,2}^{\downarrow},v_{n,1}^{\uparrow},v_{n,1}^{\downarrow},v_{n,2}^{%
\uparrow},v_{n,2}^{\downarrow}]^{T}$. The paring parameter of $i$-th tube is
defined as $\Delta(x)^{i}=g\sum_{n}(u_{n,i}^{\downarrow}v_{n,i}^{\uparrow%
\ast }f(-E_{n})+v_{n,i}^{\downarrow\ast}u_{n,i}^{\uparrow}f(E_{n}))$. $%
f(x)=1/(e^{x/k_{B}T}+1)$ is the Fermi distribution at temperature $T$. For
zero temperature, the Fermi distribution reduces to $f(x)=\Theta(-x)$ with
$\Theta$ being the heaviside function. The particle number of $i$-th tube and
spin $\sigma$ component is $n_{i}^{\sigma}=\sum_{n}(|u_{n,i}^{%
\sigma}|^{2}f(E_{n})+|v_{n,i}^{\sigma}|f(-E_{n}))$. In order to get the
information of the wave functions in real space, we expand the wave function as%
\begin{align}
u_{n,i}^{\uparrow} & =\sum_{m}A_{n,m}^{\uparrow,i}\sqrt{\frac{2}{L}}\sin(%
\frac{m\pi x}{L}),  \tag{7a} \\
u_{n,i}^{\downarrow} & =\sum_{m}A_{n,m}^{\downarrow,i}\sqrt{\frac{2}{L}}\sin(%
\frac{m\pi x}{L}),  \tag{7b} \\
v_{n,i}^{\uparrow} & =\sum_{m}B_{n,m}^{\uparrow,i}\sqrt{\frac{2}{L}}\sin(%
\frac{m\pi x}{L}),  \tag{7c} \\
v_{n,i}^{\downarrow} & =\sum_{m}B_{n,m}^{\downarrow,i}\sqrt{\frac{2}{L}}\sin(%
\frac{m\pi x}{L}).   \tag{7d}
\end{align}
In our calculation, the cutoff number of the base function is set to be $%
N_{c}=300$. The absolute value of the minimum excitation energy with different $t$ is shown in
Fig. \ref{f3}. We find that, when t is below 0.1, the absolute value of the minimum of the
Bogoliubov energy is quite small (about $10^{-4}$ ), while absolute value of the minimum
energy changes to be $\sim10^{-1}$ with tunneling above 0.1. An obvious leap
emerges at a critical value $t=0.1$. This signifies the phase transition
between tFF state and gFF state.

\begin{figure}[ptb]
\includegraphics[width=8.4cm]{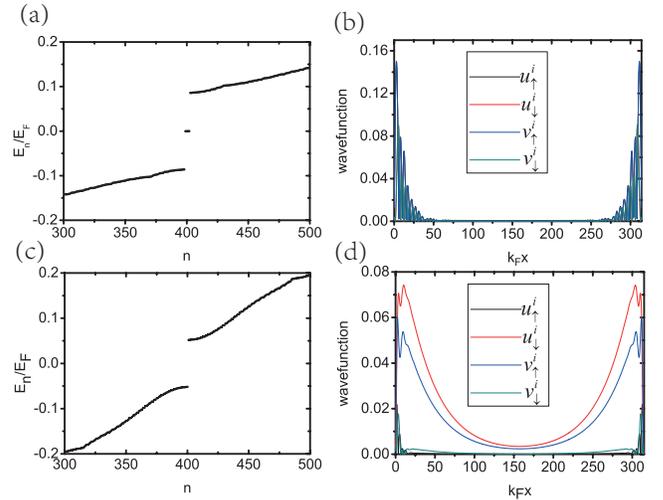}\newline
\caption{Bogoliubov spectra and wave functions with different tunneling are
shown in this figure. For our two coupled tubes model, when $t=0.05$, there
is zero mode in the spectra (in (a)), which is known as majorana fermion.
And the wave function in (b) shows edge mode in this case. In contrast, the
spectra is fully gapped if $t=0.15$. And the wave function is extensional in
(d). The other parameter of this figure is set as:$g=-0.3,h=0.2,hx=0.2,
\protect\alpha k_{F}/E_{F}=0.6$. The Fermi energy $E_{F}$ is determined by
the total particle number. }
\label{f4}
\end{figure}

Figure \ref{f4} shows the energy spectra and the wave functions of tFF state and
gFF state with different tunneling parameters. In the case of tFF phase with
$t=0.05$, there are zero modes in subgraph (a), which is known as the Majorana
fermion, and signifies the topological property in this state. The
corresponding eigen wave function in real space is also depicted in Fig.\ref%
{f4}(b), which is localized at the edge of the $\hat{x}$-direction. We note that owing
to the symmetry of tube 1 and tube 2, only the edge mode in the first tube
is shown. When $t=0.15$, which is larger than the critical point, the
spectra is fully gapped as shown in subplot (c). And the corresponding
wave function (subgraph (d)) is more extended contrast with subgraph (b).

In conclusion, we get qualitative result of the phase transition with
different tunneling. When t is small, the array of the coupling tubes is
similar to quasi one dimensional system. We show the topological phase and
topological elementary excitation. If we tune the tunneling larger, the
model is more like quasi two dimensional system. And the system is fully gapped.
With two coupled tubes model approximation, we get the critical point of the phase
transition.

\section{QUASI TWO DIMENSIONAL RESULT}

\subsection{PAIRING STATE WITH ZERO CENTER OF MASS MOMENTUM}

In last section, we analyze the phase diagrams and the topological state
with weak tunneling $t$. To get a better understanding about the crossover
physics, in this section, we focus on the quasi two dimensional geometry
where the tunneling $t$ can be changed over a wide parameter range. We
start by studying the pairing states with zero out-of-plane Zeeman field.
When $h=0$, $H_{eff}$ can be diagonalized analytically. It can be verified
numerically that in this case, paring with zero center-of-mass momentum $%
Q_{x}=0$ is stable against finite momentum paring. The analytical expression
of the quasi-particle and quasi-hole dispersions along $k_{x}=0$ axis can be
written as $E_{k_{x}=0,\pm}^{\lambda}=\lambda(\sqrt{\varepsilon_{ky}^{2}+%
\Delta^{2}}\pm h_{x})$ with $\lambda=\pm$, and $%
\varepsilon_{ky}=-2t(cos(k_{y})-1)$. Apparently, only two of the branchs $%
E_{k_{x}=0,-}^{\lambda}$ can cross zero. And this leads to the exotic nodal
superfluid (nSF) with gapless points in the excitation spectra. It is
interesting to classify nSF state with different number of gapless points.
Typically, there can be two kinds of nSF state with two or four isolated gapless
points. A simple calculation shows that when $\mu\pm \sqrt{%
h_{x}^{2}-\Delta^{2}}\in\lbrack0,4t]$, there will be four isolated gapless
points in the excitation spectra, while the excitation spectra is fully
gapped when$\mu\pm\sqrt{h_{x}^{2}-\Delta^{2}}\notin\lbrack0,4t]$.
Otherwise, we find that only two gapless points appear in the dispersion.

\begin{figure}[ptb]
\includegraphics[width=8.4cm]{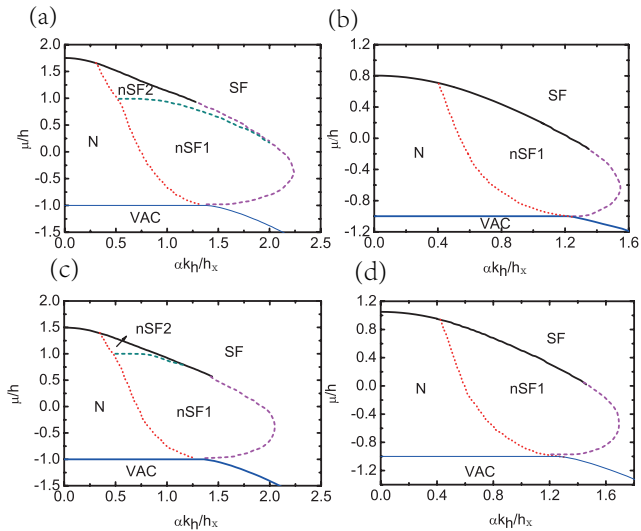}\newline
\caption{Phase diagrams with different interaction $g$ and different
tunneling $t$ on the $\protect\alpha-\protect\mu$ plane . For (a) and (b),
the parameter is set as: (a).$g=-0.35, t=1$;(b).$g=-0.65, t=1$,
respectively. For (c) and (d), the parameter is set as: (c).$g=-0.35, t=0.9$%
; (d).$g=-0.35, t=0.7$, respectively. In all the figures, the black solid
lines represent the first-order phase transition boundaries while the dashed
burgundy lines represent the continuous phase transition boundaries. The red
dotted curve corresponds to $\Delta=10^{-3}$ threshold. $h=1$ in all the
subplots.}
\label{f5}
\end{figure}

In Fig. \ref{f5}, we map out the phase diagram with different interaction $g$
on the $\alpha-\mu$ plane. Compared with phase diagrams with Rashba SOC in
two dimension \cite{Zhou}, the topological phase is now replaced by nodal
superfluid state. We focus on nodal superfluid in our quasi two dimensional
model. With anisotropic SOC, superfluid state with zero center-of mass
momentum pairing is still stable against normal state with a large in-plane
Zeeman field $h_{x}$. As we have anticipated, there exists two kinds of
nodal superfluid states with either four isolated nodes or two isolated
nodes in the excitation spectra as shown in Fig. \ref{f5}(a). The four-node
superfluid disappears in Fig. \ref{f5}(b) as we increase the interaction
strength to $g=-0.65$. It means that the nodal superfluid is more stable with
weaker interaction. To study the phase transition in the case of dimensional
crossover from 1D to 2D, we also plot the phase diagrams with different
tunneling $t$ as shown in subgraphs (a,c,d) of Fig. \ref{f5}. For fixed
interaction strength $g=-0.35$, one can see that the four-nodal superfluid
region shrinks rapidly and finally disappears from the phase diagram with $%
t=0.7$ (Fig. 5d). Therefore, the four-nodal phase is more stable with
relatively larger ratio $|t/g|$. We note that in these cases, nodal phases
process a relatively large region. This is very different from the
quasi one dimensional case, where gFF state dominates the superfluid region,
and tFF state disappears very quickly as we increase the inter-tube
tunneling. Therefore, we can claim that nodal-superfluid is more stable in
quasi two dimensional geometry than the quasi one dimensional case.

\begin{figure}[ptb]
\includegraphics[width=8.7cm]{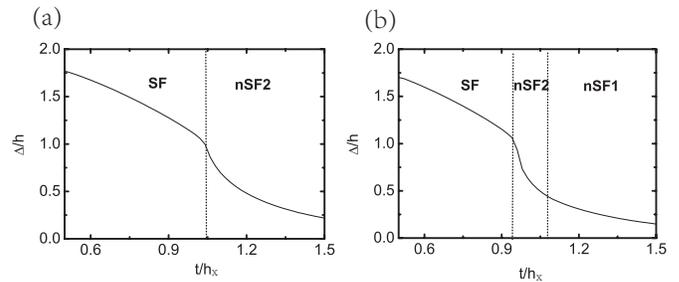}\newline
\caption{Phase diagrams with different tunneling $t$. The chemical potential
$\protect\mu$ in (a) and (b) is set as $\protect\mu=0.6$ and $\protect\mu=0.9
$, respectively. The other parameters are set as follows: $g=-0.3;\protect%
\lambda k_{h}/h_{x}=1.2,h_{x}=1,h=0$. }
\label{f6}
\end{figure}

To illustrate the crossover behaviors of the SF phases along with the
tunneling, we also plot the evolution of pairing parameter $\Delta$ as a
function of $t/h_{x}$ from 0.5 to 1.5 by fixing all other parameters. As
shown in Fig. \ref{f6}(a), one can see below a critical tunneling value $%
t_{c}/h_{x}=1.04$, $\Delta$ changes smoothly. This corresponds to the normal
paring state and the excitation spectra is gapped. However, the gap closes and the
ground state changes to a four-node paring phase (nSF2) when the tunneling
is above the critical value. This transition is of first-order which can be
verified by the the discontinuity of $\partial\Delta/\partial t$ across the
transition point. The ground state becomes more complicated in Fig.\ref{f6}%
(b) where two critical values $t_{c1}/h_{x}=0.94$ and $t_{c2}/h_{x}=1.08$
are involved. Besides a first-order transition from the normal paring state
to the nSF2 phase, the ground state changes to nSF1 state when the
tunneling is above $t_{c2}/h_{x}$. We note that in this case, the
transition is continuous which is also consistent with the results shown in Fig. %
\ref{f5}.

\subsection{FF PARING STATE WITH OUT-OF-PLANE ZEEMAN FIELD}

When transverse Zeeman field is introduced, the presence of SOC leads to a
asymmetric deformed Fermi surface. In this case, s-wave paring with zero
center-of mass momentum becomes unfavorable, which is replaced with FF
states with finite $Q_{x}$. Qualitatively, this corresponds to a shift
about the local minimum of the thermodynamic potential from $Q_{x}=0$ to a
finite $Q_{x}$ plane. When the shift is small, most of the properties of the
FF state are similar with the paring states with zero center-of-mass
momentum.

\begin{figure}[ptb]
\includegraphics[width=8.9cm]{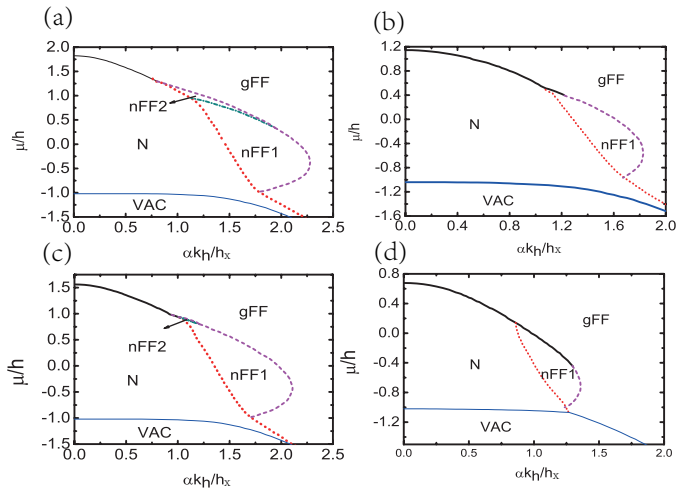}\newline
\caption{Phase diagram of finite momentum paring state in $\protect\alpha-%
\protect\mu$ plane. We still analyze quantum phase with different
interaction and different tunneling. For (a) and (b), the parameter is set
as: (a).$g=-0.35$;(b).$g=-0.65$, respectively. The other parameter for the
two figures are set as: $t=1, h_{x}=1,h=0.2$. For (c) and (d), the parameter
is set as: (c).$t=0.9$; (d).$t=0.5$, respectively. The other parameters of
the two figures are set as: $g=0.35, h_{x}=1;h=0.2$. In all the figures, the
black solid lines represent the first-order phase transition boundaries
while the dashed burgundy lines represent the continuous phase transition
boundaries. The red dotted curve corresponds to $\Delta=10^{-3}$ threshold. }
\label{f7}
\end{figure}

In Fig.\ref{f7}, we map out the phase diagrams in the $\alpha-\mu$ plane
with different interaction and different tunneling. The calculation shows
that these diagrams share similar patterns shown in Fig. \ref{f5}, except
that the nodal superfluid states are now replaced by nodal FF states. We
stress that in all these diagrams, FF state is hosted in the $\alpha-\mu$
plane. This is very different from the usual two dimensional polarized Fermi
gases without SOC, where FFLO states can only be stabilized within a small
parameter region. Therefore, FF states becomes more stable due to the
presence of transverse Zeeman field and SOC, which greatly increases the
possibility to find such exotic pairing states in our system.

Fig. \ref{f7}(a) and \ref{f7}(b) show the phase diagrams with different
interactions. We can find the first order phase transition line is longer
with larger interaction. In addition, nodal FF state is more favourable with
smaller interaction. For instance, when interaction strength $g=-0.35$,
there are two kinds of nodal-FF state, two-node FF state (nFF1) and
four-node FF state (nFF2). As the interaction strength $g$ increases, nFF2
disappears gradually and the regime for nFF1 phase also becomes smaller. It
is expected that when the interaction is very strong, all the nodal FF
states will be replaced by fully gapped FF states.

\begin{figure}[ptb]
\includegraphics[width=8.7cm]{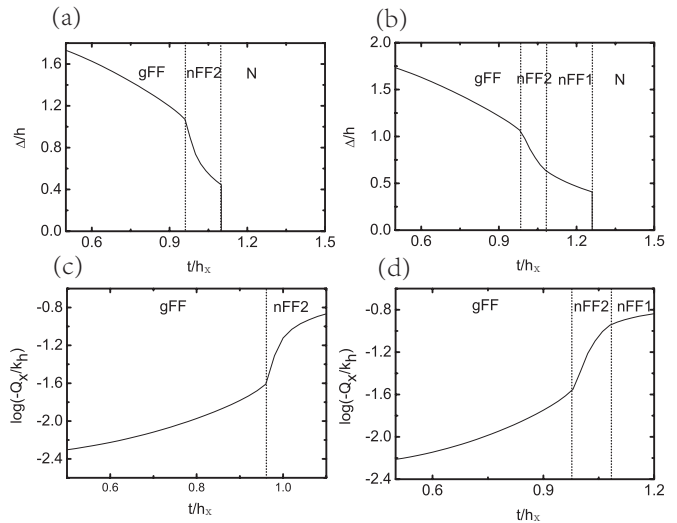}\newline
\caption{Phase diagrams of the the paring order parameter $\protect\delta$
and center-of-mass momentum $Q_{x}$ with different tunneling $t$. The
chemical potential $\protect\mu$ and SOC strength in (a)/(c) and (b)/(d) is
set as $\protect\mu=0.6, \protect\lambda k_{h}/h_{x}=1.25$ and $\protect\mu%
=0.9, \protect\lambda k_{h}/h_{x}=1.5$, respectively. The other parameters
are set as follows: $g=-0.35, h_{x}=1,h=0.2$. }
\label{f8}
\end{figure}

To illustrate the crossover physics of FF states, we also plot the phase
diagrams for different tunnelings $t=(1,0.9,0.5)$, as shown in Fig. (\ref%
{f7}a), (\ref{f7}c) and (\ref{f7}d) respectively. When $t=(1,0.9)$, both
nFF1 and nFF2 phase appear in the phase diagram, although the region of nFF2
shrinks rapidly as $t$ decreases. When the tunneling decreases to 0.5, only
nFF1 is left in the phase diagram. Therefore, we assert that nFF2 is
favourable with larger tunneling. In this case, the system is more
appropriate to be treated as a two dimensional one. When tunneling decrease
further, the situation becomes more complicated, and tFF state comes into
play as shown in Fig. \ref{f2}. Fig. \ref{f8} shows the changes of the
pairing order parameter and center-of-mass momentum as we increase the
inter-tube tunneling $t$. One can see that fully gapped FF state always
exists in the dimensional crossover region, while nodal FF state emerges
around $t=1$. The center-of-mass momentum $|Q_{x}|$ increases almost
monotonically along with $t$. The ground state changes to normal state when $%
t$ is larger enough.

\section{CONCLUSION}

In summary, we have studied the paring state in the dimensional crossover
region in an array of coupling tubes. The interplay of SOC, Zeeman field,
and tunneling lead to rich phase structure with exotic pairing states. In
particular, the ground state is FFLO state with finite center-of-mass
momentum pairing over large parameter region, which facilitates the
experimental verification of this kind of exotic paring state. For quasi
one dimensional system, there is tFF state in the phase diagram, which is
characterized by the presence of Majorana edge state in the model. In order
to qualitatively show the transition from the tFF state to normal SF state
in the crossover region, a toy model composed of two-coupled tubes is
studied to show the critical tunneling. Further more, the phase diagrams
with different interaction and inter-tube tunneling are also presented.
Except for fully gapped states, we find that there is a large parameter
region for which pairing states with isolated gapless excitations in the
dispersion appears. nSF and nFF states exist when the system is
a quasi two dimensional system, and finally they change to normal state with further
increasing the inter-tube tunneling.

\textbf{Acknowledgments:} We thank Wei Yi, Fan Wu, Bei-Bin Huang for helpful
discussions. This work is supported by NSFC under Grant No. 11504038, "
Chongqing Fundamental, Frontier Research Program" (Grant No.
cstc2015jcyjA00013), and Foundation of Education Committees of Chongqing
(Grant No. KJ1500411). X.-F Zhou acknowledges the support from the National
Fundamental Research Program of China (Grants No. 2011CB921200 and No.
2011CBA00200), the Strategic Priority Research Program of the Chinese
Academy of Sciences (Grant No. XDB01030200), and NSFC (Grant Nos. 11474266).
L.Wen acknowledges the support from NSFC under Grant No. 11504037, "
Chongqing Fundamental and Frontier Research Program" (Grant No.
cstc2015jcyjA50024), and Foundation of Education Committees of Chongqing
(Grant No. KJ1500311).




\begin{thebibliography}{99}
\bibitem{niu} D.Culcer, J.Sinova, N.A.Sinitsyn, T.Jungwirth, A.H.MacDonald,
Q.Niu, Phys. Rev. Lett. $\mathbf{93}$, 046602(2004)

\bibitem{xie} Q.-F.Sun, X.-C.Xie, Phys. Rev. B, $\mathbf{71}$, 155321(2005)

\bibitem{Hsieh} D.Hsieh, F.Mahmood, J.W.Mclver, D.R.Gardner, Y.S.Lee, N.Gedik, Phys. Rev. Lett, $\mathbf{107}$, 077401(2011)

\bibitem{kane1} M.Z.Hasan, C.L.Kane, Rev. Mod. Phys, $\mathbf{82}$,
3045(2010)

\bibitem{Qi} X.-L.Qi, S.-C.Zhang, Rev. Mod. Phys. $\mathbf{83}$, 1057(2011)

\bibitem{kane} C.L.Kane, E.J.Mele, Phys. Rev. Lett. 95, 146802(2005)

\bibitem{Lin} Y.-J.Lin, K.Jimenez-Garcia, and I.B.Spielman, Nature $\mathbf{%
471}$, 83(2011)

\bibitem{Wang} P.Wang, Z.-Q.Yu, Z.Fu, J.Miao, L.Huang, S.Chai, H.Zhai, and
J.Zhang, Phys.Rev.Lett.$\mathbf{109}$,095301(2012)

\bibitem{Zwierlein} L.W.Cheuk, A.T.Sommer, Z.Hadzibabic, T.Yefsah, W.S.Bakr,
and M.W.Zwierlein, Phys.Rev.Lett.$\mathbf{109}$,095302(2012)

\bibitem{zhang1} P.-J Wang, Z.-Q Yu, Z.-K Fu, J.Miao, L.-H Huang, S.-J Chai,
H.Zhai, J.Zhang, Phys. Rev. Lett. $\mathbf{109}$, 095301 (2012)

\bibitem{zhang2} Z.-K Fu, L.-H Huang, Z.-M Meng, P.-J Wang, X.-J Liu, H.Pu,
H.Hu, J.Zhang, Phys. Rev. A $\mathbf{87}$, 053619(2013)

\bibitem{zhang3} L.-H Huang, P.-J Wang, P.Peng, Z.-M Meng, L.-C Chen,
P.Zhang, J.Zhang, arXiv:1504.02021(2015)

\bibitem{zhang4} L.-H Huang, Z.-M Meng, P.-J Wang, P.Peng, S.-L Zhang, L.-C
Chen, D.H Li, Q.Zhou, J.Zhang, arXiv:1506.02861(2015)

\bibitem{Yu} Z.-Q Yu, H.Zhai, Phys. Rev. Lett. $\mathbf{107}$, 195305(2011)

\bibitem{Gong} M.Gong, S.Tewari, and C.-W Zhang, Phys. Rev. Lett. $\mathbf{%
107}$, 195303(2011)

\bibitem{Zhou} J.Zhou, W.Zhang, W.Yi, Phys. Rev. A $\mathbf{84}$,
063603(2011)

\bibitem{Cheng} G.Chen, M.Gong, and C.-W Zhang, Phys. Rev. A. $\mathbf{85}$%
, 013601(2012)

\bibitem{He1} L.-Y He, X.-G Huang, Phys. Rev. Lett. $\mathbf{108}$, 145302
(2012)

\bibitem{He2} L.-Y He, X.-G Huang, Phys. Rev. B. $\mathbf{86}$, 014511 (2012)

\bibitem{He3} L.-Y He, X.-G Huang, H. Hu, X.-J Liu, Phys. Rev. A. $\mathbf{87%
}$, 053616(2013)

\bibitem{Shenoy} J.P.Vyasanakere, S.-Z Zhang, V.B.Shenoy, Phys. Rev. B. $%
\mathbf{84}$, 014512(2011)

\bibitem{Gong2} M.Gong, G.Chen, S.Jia, C.Zhang, Phys. Rev. Lett. $\mathbf{109%
}$, 105302(2012)

\bibitem{Hu0} X.-J Liu, H.Hu, Phys. Rev. A. $\mathbf{88}$, 023622(2013)

\bibitem{Hu1} H.Hu, L.Jiang, H.Pu, Y.Chen, X.-J Liu, Phys. Rev. Lett. $\mathbf{%
110}$, 020401(2013)

\bibitem{Hu2} S.-G Peng, X.-J Liu, H.Hu, K.-J Jiang, Phys. Rev. A. $\mathbf{%
86}$, 063610(2012)

\bibitem{Hu3} H.Hu, H.Pu, J.Zhang, S.-G Peng, X.-J Liu, Phys. Rev. A. $%
\mathbf{86}$, 053627(2012)

\bibitem{Hu4} X.-J Liu, H.Hu, Phys. Rev. A. $\mathbf{85}$, 033622(2012)

\bibitem{Hu5} X.-J Liu, L.Jiang, H.Pu, Hui Hu, Phys. Rev. A. $\mathbf{85}$,
021603(R)(2012)

\bibitem{Hu6} L.Jiang, X.-J Liu, H.Hu, H.Pu, Phys. Rev. A. $\mathbf{84}$,
063618(2011) 

\bibitem{higgs1} M.Endres, T.Fukuhara, D.Pekker, M.Cheneau, P.Schau$\beta$,
I.Bloch, Nature, $\mathbf{487}$, 454(2012)

\bibitem{higgs2} L.Pollet, N.V.Prokofev, Phys. Rev. Lett. $\mathbf{109}$,
010401(2012)

\bibitem{bloch2} I.Bloch, J.Dalibard, W.Zwerger, Rev. Mod. Phys. $\mathbf{80}
$, 885(2008)

\bibitem{duan} L.-M.Duan, E.Demler, M.D.Lukin, Phys. Rev. Lett. $\mathbf{91}$%
, 090402(2003)

\bibitem{FF1} P.Fulde and R.A.Ferrell, Phys. Rev. $\mathbf{135}$, A550(1964)

\bibitem{FF2} A.I.Larkin and Y.N.Ovchinnikov, Zh. Eksp. Teor. Fiz. $\mathbf{%
47}$, 1136(1964) 

\bibitem{wu1} H.A.Radovan et.al., Nature, $\mathbf{425}$, 51(2003)

\bibitem{wu2} A.Bianchi, R.Movshovich, C.Capan, P.G.Pagliuso, and J.L.Sarrao, Phys. Rev. Lett. $\mathbf{91}$, 187004(2003)

\bibitem{wu3} K. Kakuyanagi, et al., Phys. Rev. Lett. $\mathbf{94}$
047602(2005)

\bibitem{org1} S.Uji, H.Shinagawa, T.Terashima, T.Yakabe, et.al., Nature $%
\mathbf{410}$, 908(2001).

\bibitem{fflo1} Y.Liao, A.S.C.Rittner, T.Paprotta, W.-H.Li, G.B.Partridge,
R.G.Hulet, S.K.Baur and E.J.Mueller, Nature $\mathbf{467}$, 567(2010).

\bibitem{fflo2} W.Ketterle and M.W.Zwierlein, arXiv.org:0801.2500

\bibitem{fflo3} Y.L.Loh and N.Trivedi, Phys. Rev. Lett. $\mathbf{104}$,
165302(2010)

\bibitem{fflo4} M.M.Parish, S.K.Baur, E.J.Mueller, and D.A.Huse, Phys. Rev.
Lett. $\mathbf{99}$, 250403(2007).

\bibitem{fflo5} J.M.Edge and N.R.Cooper, Phys. Rev. Lett. $\mathbf{103}$,
065301(2009).

\bibitem{fflo6} K.Yang, Phys. Rev. Lett. $\mathbf{95}$, 218903(2005).

\bibitem{Qu} C.-L Qu, Z.Zheng, M.Gong, et.al., Nature Communication $\mathbf{%
4}$, 2710(2013)

\bibitem{Yi} W.Zhang, W.Yi, Nature Communication $\mathbf{4}$, 2711 (2013)

\bibitem{Liu} X.-J Liu, H.Hu, Phys. Rev. A $\mathbf{88}$, 023622 (2013)

\bibitem{Cao} Y.Cao, S.-H Zou, X.-J Liu, et.al., Phys. Rev. Lett. $\mathbf{%
113}$, 115302(2014)

\bibitem{Sun} K.Sun, C.J. Bolech, Phys. Rev. A $\mathbf{87}$, 053622 (2013)

\bibitem{Zongshu} M.Z.Hasan and C.L.Kane, Rev. Mod. Phys. 82, 3045 (2010)

\bibitem{Duan} L.-M.Duan, E.Demler, M.D.Lukin, Phys. Rev. Lett. $\mathbf{91}$%
, 090402(2003)

\bibitem{Pf} A.Y.Kitaev, Ann. Phys. $\mathbf{321}$, 2111(2003)
\end{thebibliography}
\end{document}